# PixCUE: Joint Uncertainty Estimation and Image Reconstruction in MRI using Deep Pixel Classification

Mevan Ekanayake[a,b,^], Kamlesh Pawar[a,^], Gary Egan[a,c], and Zhaolin Chen[a,d,*]

[a]Monash Biomedical Imaging, Monash University, Clayton, VIC 3800 Australia
[b]Department of Electrical and Computer Systems Engineering, Monash University, Clayton, VIC 3800 Australia
[c]School of Psychological Sciences, Monash University, Clayton, VIC 3800 Australia
[d]Department of Data Science and AI, Monash University, Clayton, VIC 3800 Australia

## Abstract

Deep learning (DL) models are capable of successfully exploiting latent representations in MR data and have become state-of-the-art for accelerated MRI reconstruction. However, undersampling the measurements in k-space as well as the over- or under-parameterized and non-transparent nature of DL make these models exposed to uncertainty. Consequently, uncertainty estimation has become a major issue in DL MRI reconstruction. To estimate uncertainty, Monte Carlo (MC) inference techniques have become a common practice where multiple reconstructions are utilized to compute the variance in reconstruction as a measurement of uncertainty. However, these methods demand high computational costs as they require multiple inferences through the DL model. To this end, we introduce a method to estimate uncertainty during MRI reconstruction using a pixel classification framework. The proposed method, PixCUE (*stands for Pixel Classification Uncertainty Estimation*) produces the reconstructed image along with an uncertainty map during a single forward pass through the DL model. We demonstrate that this approach generates uncertainty maps that highly correlate with the reconstruction errors with respect to various MR imaging sequences and under numerous adversarial conditions. We also show that the estimated uncertainties are correlated to that of the conventional MC method. We further provide an empirical relationship between the uncertainty estimations using PixCUE and well-established reconstruction metrics such as NMSE, PSNR, and SSIM. We conclude that PixCUE is capable of reliably estimating the uncertainty in MRI reconstruction with a minimum additional computational cost.

**Keywords:** MR image reconstruction, deep learning uncertainty estimation, convolutional neural network, pixel classification framework

___________

^These authors contributed equally
*Corresponding author. *Tel.*: +61 3 9905 0841; *E-mail address*: zhaolin.chen@monash.edu

# 1 Introduction

Image reconstruction in MRI involves processing the raw k-space data acquired during a scan to human-readable image formats. In brief, the process involves applying multi-dimensional Fourier transform to the raw k-space data and combining the images from multiple channels [1]. However, to reduce the scan time of MR imaging, undersampling in the k-space is often performed, and thus more sophisticated methods such as parallel imaging [2]–[5], compressed sensing [6]–[9], and deep learning (DL) [10]–[17] are needed to reconstruct an image with minimal undersampling artifacts. Mathematically, these methods are essentially solving an inverse problem [15], [18], [19]. This inverse problem can be solved using an explicit matrix inversion [3], [4] in parallel MRI, or by using iterative gradient descent-based methods [20]–[24] in compressed sensing. In the DL-based approaches, such inverse problems are approximated by a multi-layered model with the parameters learned from a training dataset. This approach of using training data to find a model of image reconstruction [10]–[12], [25]–[28] has been demonstrated to outperform the conventional methods of compressed sensing and parallel imaging.

Although the DL-based image reconstruction methods demonstrate significant improvements, the generalization [29]–[32] of these methods is still an area that is poorly exploited. When a model is trained on a particular set of data and is inferred on a different set of data, it may create false features that are not present in the original image (hallucinations) [32], [33]. In addition, the model may become unstable [34] as a result of a change in the input data that is not noticeable. This aspect of DL models has motivated researchers to search for methods to detect model failure modes that will alleviate potential misdiagnoses due to errors in DL image reconstruction.

Uncertainty modeling [35]–[38] is one method to detect failure modes. Inspired by computer vision research [35], uncertainty modeling has been explored in the context of MRI [37]–[39]. Specifically, in literature, the variance among multiple test inferences using Monte Carlo (MC) sampling with probabilistic models is a surrogate estimate of uncertainty [40]. Several other methods involve learning the entire system parameters utilizing invertible neural networks [41], [42]. Also, Bayesian inference methods have been introduced for uncertainty estimation [38]. Stein's Unbiased Risk Estimator (SURE) has also been utilized as a method of uncertainty estimation [43]. In several previous works, a separate uncertainty layer [37], [44] or a model [38], [45] have been utilized to estimate uncertainty. More recent comprehensive work on DL uncertainty estimation involves leveraging variational autoencoders (VAEs) to develop probabilistic reconstructions that encode the acquisition uncertainty in the latent space [37]. This approach is capable of developing a posterior to the image data from which a variance map can be generated utilizing MC sampling. The mainstream drawback of MC sampling-based uncertainty estimation specifically in the DL setting is the high computational cost during multiple inferences through large DL models.

In this study, we propose an uncertainty estimation method referred to as PixCUE (*stands for Pixel Classification Uncertainty Estimation*) that jointly performs image reconstruction and uncertainty estimation simultaneously using a pixel classification framework. PixCUE produces the reconstructed image along with an uncertainty map during a single forward pass through the DL model. The pixel classification framework here converts the reconstruction problem to a SoftMax classification problem where the predicted probability distribution of each pixel is utilized to estimate its uncertainty. We observe that PixCUE is capable of generating uncertainty maps that highly correlate with the actual error in reconstruction with respect to various MR imaging sequences and under numerous adversarial conditions. We also observe that the uncertainty estimations obtained using PixCUE are highly correlated with that of the conventional MC method, yet substantially faster compared with the MC method. We further provide an empirical relationship between uncertainty estimations using PixCUE and well-established reconstruction metrics such as NMSE, PSNR, and SSIM.

The remainder of the paper is arranged as follows. Section 2 provides the background to the underlying work in this paper. Section 3 presents the proposed uncertainty estimation method. In Sections 4 and 5, we present our experiments and the results, respectively. A comprehensive discussion and a conclusion are presented in Sections 6 and 7, respectively.

## 2 Background

### 2.1 Fundamentals of MRI Reconstruction

In MRI, the acquired k-space data is represented as:

$$y = Fx + \eta \tag{1}$$

where $y$ is the k-space data, $x$ is the complex-valued image, $F$ is the Fourier transform operator, and $\eta$ represents tissue and instrumentation noise. To reduce the scan time, the k-space data is often undersampled using a predefined undersampling pattern. Eq. 1 can be rewritten considering the undersampling as:

$$y_u = MFx + \eta' = F_u x + \eta' \tag{2}$$

where $y_u$ is the undersampled k-space data, $M$ is the undersampling operator, $\eta'$ represents noise, and $F_u$ represents the partial Fourier transform operator accounted for undersampling. The task in MRI reconstruction is to recover the MRI image with diagnostic quality from $y_u$. Conventional solutions for MRI reconstruction utilized linear analytic techniques like partial Fourier encoding, sensitivity encoding (SENSE) [4], simultaneous acquisition of spatial harmonics (SMASH) [46], and generalized auto-calibrating partially parallel acquisitions (GRAPPA) [3]. These methods utilized knowledge of the imaging systems and k-space correlations. Subsequently, the reconstruction methods shifted towards non-linear iterative algorithms that incorporated the physics of the imaging system and regularization of priors such as sparsity [47], artificial sparsity [48], [49], low rank [50], [51], manifold [52], total variation [53], and dictionary learning [54].

### 2.2 Deep Learning MRI Reconstruction

In the context of DL, the image reconstruction from undersampled k-space data is conventionally modeled as a regression problem using a DL network, $G_\theta$ parameterized by $\theta$ as:

$$x^* = G_\theta(F^H y_u) \tag{3}$$

where $x^*$ is the reconstructed image using a DL network and $F^H$ represents inverse Fourier transform operator. Different manifestations of $G_\theta$ and the loss functions to train such a network have been reported in the literature [11], [55], [56]. Most of the DL-based MRI reconstruction methods utilize convolutional neural networks (CNN) and can be summarized under two main groups [57]: data-driven end-to-end DL methods that map low-quality undersampled images to high-quality references [12], [16], [33], [58], and physics-constrained DL methods that iteratively solve inverse problems [15], [59], [60]. Recently, there have been several works involving transformer networks which attempt to aggregate global correlations among MR image patches [61], [62].

**2.3 Uncertainty Estimation in Deep Learning MRI Reconstruction**

The robustness of DL networks for inverse problems such as MRI reconstruction has not been thoroughly investigated, and hence, prone to introduce image artifacts, especially when subjected to out-of-distribution inputs. Also, there is currently a lack of generalizable methods for estimating the uncertainty in DL reconstructions [37]. Reliable uncertainty estimation methods could be useful both as an evaluation metric and for gaining the interpretability of a given reconstruction model or a dataset [35]. Radiologists could utilize uncertainty maps in combination with the reconstructed MR image to make better-informed judgments.

To this end, several uncertainty estimation strategies have been proposed in the context of DL reconstruction. Edupuganti et al. [37] utilized VAEs to create a probabilistic reconstruction approach, which utilizes MC sampling to generate uncertainty from the image's posterior. Kitichotkul et al. [43] utilized a CNN-based SURE framework to create heatmaps as per-pixel confidence intervals for compressed sensing MRI which communicated the reliability of reconstruction to the end-users. Zhang et al. [39] introduced an MRI reconstruction method that selects measurements dynamically and iteratively to reduce the uncertainty in reconstruction. Narnhofer et al. [63] proposed a deterministic MRI reconstruction that employs a Bayesian framework for uncertainty quantification in single and multi-coil undersampled MRI reconstruction. More recent work on Diffusion probabilistic Models [64] suggests drawing samples from the posterior distribution given the measured k-space using the Markov chain Monte Carlo (MCMC) method and computing uncertainty maps from those drawn samples.

**2.4 Monte Carlo Drop-Outs for Uncertainty Estimation in Deep Learning MRI Reconstruction**

Although, conventionally DL MRI reconstruction problem is modeled using a fixed set of model parameters, in the Bayesian Neural Network (BNN) approach, a distribution over the model parameters is learned. Let $D_{tr} = \{X, Y\}$ denote a training dataset where $\{X, Y\}$ are input-output paired data. In BNN formulation, the output for an arbitrary sample $X^*$ from the test dataset, can be predicted with respect to the posterior distribution, $p(\theta|X, Y)$ as below [65]:

$$p(Y^*|X^*, D_{tr}) = \int p(Y^*|X^*, \theta) p(\theta|D_{tr}) d\theta \qquad (4)$$

where $p(\theta|X, Y) = \frac{p(Y|X, \theta) p(\theta)}{p(Y|X)}$ and is intractable, i.e., cannot compute analytically. However, $p(\theta|X, Y)$ can be approximated using another distribution $q(\theta)$ whose structure is easy to evaluate. This can be done through variational inference techniques [66], [67], [67] and the approximation is usually performed by minimizing Kullback-Leibler (KL) divergence [68] between the variational inference and the posterior distribution so that $q(\theta)$ is as close as possible to $p(\theta|X, Y)$.

In the MC Drop-Out approach, $q(\theta)$ is parametrized using the drop-out of the model weights [69], i.e., $q_\alpha(\theta)$ where $\alpha$ denotes the drop-out fraction. In this approach, $q_\alpha(\theta)$ is simply enforced over the model weights before each layer of the DL model. During inference, uncertainty can be computed using multiple forward passes with different drop-out realizations which is known as an MC simulation. By performing $T$ stochastic forward passes and computing the variance across all the generated $T$ outputs, the uncertainty can be formulated as the predictive variance of the output reconstruction estimation, $\hat{x}$:

$$U_m = \text{Var}(\hat{x}) = \frac{1}{T} \sum_{t=1}^{T} \left( \hat{x}_t - \frac{1}{T} \sum_{t=1}^{T} \hat{x}_t \right)^2 \qquad (5)$$

where $U_m$ denotes the uncertainty estimation using the MC Drop-Outs approach and $\hat{x}_t$ is the reconstructed image during the $t^{\text{th}}$ forward pass.

# 3 Methods

## 3.1 Pixel Classification framework for MRI Reconstruction

Our previous work [16] demonstrated that an image reconstruction task can be transformed into a classification task by the quantization of the target image. To transform the image reconstruction task into a pixel classification task, first, the target image is converted to an *n-bit* unsigned representation where each pixel in the target image can assume to take only one of the $2^n$ distinct pixel intensity levels. Since there are only $2^n$ distinct pixel intensity levels, we can design a DL network that can classify each pixel into one of these $2^n$ classes. Therefore, the output of the DL pixel classification network, $x' = G'_\theta(F^H y_u)$ is of size $N \times N \times D$, where $D = 2^n$ is the number of classes (or pixel intensity levels) and N is the spatial dimension of the image. $G'_\theta$ is a DL pixel classification network parameterized by $\theta$. The last layer of the network consists of a softmax function along the class dimension that makes the sum of predicted output equals one along the class dimension, i.e. $\sum_{c=0}^{D-1} x'_r(c) = 1$; $\forall r$, where $x'_r$ is the output at pixel location, $r$, resulting in a predicted probability distribution for each pixel. The network $G'_\theta$ can be trained with the categorical cross-entropy loss, where the loss at pixel location $r$ is given by:

$$\mathcal{L}_r = \sum_{c=0}^{D-1} -x_r^{tar}(c) \log x'_r(c) \qquad (6)$$

where $x_r^{tar}(c) = \begin{cases} 1 \text{ if } c = h_r \\ 0 \text{ if } c \neq h_r \end{cases}$ is the target which constitutes a one-hot encoding of the true labels for the classification where $h_r$ represents the pixel intensity level at pixel location $r$ of the quantized target image. During the training of the pixel classification network, $x'_r$ converges to a vector that contains the predicted probability values for the corresponding quantized intensity levels. The final image $\hat{x}$ can be reconstructed in a pixel-wise manner by taking a weighted average of intensity levels (weighted by their corresponding predicted probabilities), and normalizing it as below:

$$\hat{x}_r = \frac{1}{D-1} \sum_{c=0}^{D-1} c \, x'_r(c) \tag{7}$$

where $\hat{x}_r$ is the final image value at pixel location $r$. It should be noted that $x'$ is only utilized to train the network but the pixels in the final image were reconstructed using Eq. 7 resulting in floating point output values.

### 3.2 Relationship between Uncertainty and Variance

The variables $x'_r$ and $x_r^{tar}$ can be considered as predicted and target probability distributions, respectively for each pixel. As shown in Eq. 6, we minimize the categorical cross-entropy between $x_r^{tar}$ and $x'_r$. Since $x_r^{tar}$ contains only a single non-zero entry, it can be interpreted as a Dirac delta function [70] and is constant for a given pixel making the entropy, $H(x_r^{tar}) = 0$. Thus, the categorical cross-entropy between $x_r^{tar}$ and $x'_r$ becomes equivalent to the KL-divergence between $x_r^{tar}$ and $x'_r$ given by:

$$D_{KL}(x_r^{tar} || x'_r) = \sum_{c=0}^{D-1} x_r^{tar}(c) \log\left(\frac{x_r^{tar}(c)}{x'_r(c)}\right) \tag{8}$$

Therefore, the cross-entropy minimization in Eq. 6 results in minimizing $D_{KL}(x_r^{tar} || x'_r)$ in Eq. 8 and the optimal solution is obtained when $x'_r(c) = x_r^{tar}(c)$. However, in reality $x'_r(c)$ will only approximate $x_r^{tar}(c)$ due to the variability in the input datasets (e.g. noise, MRI artifacts, and pathology introduced out-of-distribution samples) as well as the imperfect modeling process (e.g. over- or under-fitted models), thus uncertainty is introduced. Such uncertainty is manifested as variance in the predicted probability distribution $x'_r(c)$ [69].

### 3.3 Pixel Classification Uncertainty Estimation (PixCUE)

We implement the fastMRI variational network (VN) architecture [27] as the base network of PixCUE. The VN architecture consists of unrolled CNNs with data consistency layers and at each iteration the k-space data ($y'_k$) is modified as follows:

$$y'_{k+1} = y'_k - \alpha_k M(y'_k - y_u) + FEG_k(RF^H y'_k) \tag{9}$$

where $G_k$ is the DL network at $k^{th}$ iteration. For multichannel k-space data, we utilized two operators: (i) a *Reduce (R)* operator which combines the multichannel images to a single channel, and (ii) an *Expand (E)* operator which creates a multichannel image from the single channel image using estimated sensitivity maps. Other notations are described as previously. The estimation of the sensitivity maps is also learned during the training using a separate sensitivity estimation DL network. Interested readers can refer to the fastMRI VN paper [27] for a detailed description of the architecture.

To implement the pixel classification network, only the last iteration of the VN architecture was modified such that the output from the VN is of size $N \times N \times D$, where $D = 2^n$ is the number of classes. For training, the loss was calculated on the output probabilities as per Eq. 6 and the reference image was quantized to 8-bit ($D = 2^8$ pixel intensity levels). The trained classification network predicts the probability distribution for each pixel, $x'_r$ as seen in Fig. 1, which then allows calculating the variance of the distribution. The proposed PixCUE uncertainty estimation can be computed as below:

$$U_{p,r} = \frac{1}{D-1} Var(x'_r) \tag{10}$$

where $U_{p,r}$ is the normalized PixCUE uncertainty estimation at location $r$ and $Var(x'_r)$ is simply the variance of the predicted probability distribution. It should be noted that, unlike split prediction head methods in the literature [44], in our proposed PixCUE method, the variance of prediction distributions emerges naturally from the pixel classification approach rather explicitly enforced as a result of minimizing the cross entropy. For regions of the reconstructed image where the uncertainty is high, the network's predicted probability $x'_r$ is low. For regions where the uncertainty is low (such as the background in the image), the predicted probability $x'_r$ is high and is a close approximation of the Dirac delta function.

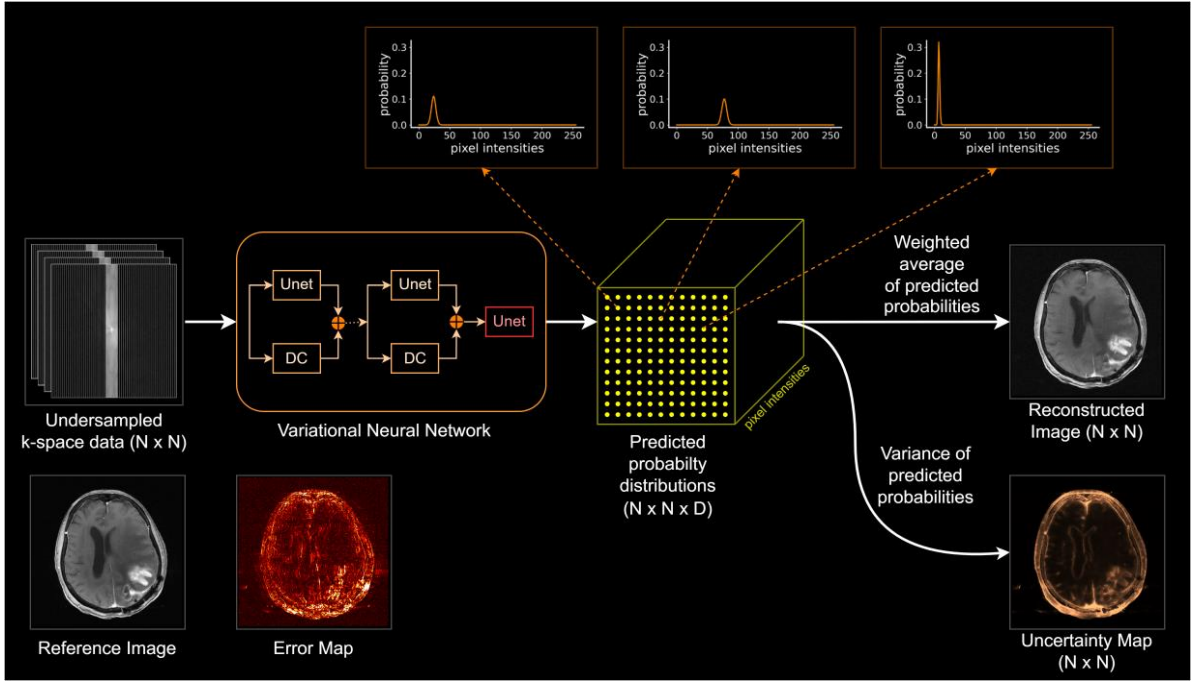

Fig. 1: Overview of the PixCUE framework for uncertainty estimation in MRI Reconstruction.

### 3.4 Variance Computation

The proposed PixCUE framework has a unique characteristic which other conventional classification tasks in computer vision do not exhibit. The predicted distribution at each pixel location demonstrates a Gaussian shape and is symmetrical about the mean as can be seen in Fig. 1. This Gaussian profile arises as a result of the sorted ordering of class labels (or pixel intensity levels) from 0 to $D-1$ which represents the tissue contrast. Hence, the predicted probability distribution can be fitted by a standard normal distribution, $x'_r \sim N(\hat{x}_r, \sigma_r^2)$ with the mean inherently being $\hat{x}_r$ (the weighted average pixel intensity computed using Eq. 7) and the variance $\sigma_r^2$ being the uncertainty in prediction. In other words, the shape of the predicted probability distribution is broader for the regions of high uncertainty and narrower for the regions with low uncertainty.

There are a couple of ways to compute the variance of $\hat{x}_r$ given its unique characteristics. The most accurate way is to fit $x'_r$ with a Gaussian function that has a mean of $\hat{x}_r$ and optimizing its variance. However, fitting a Gaussian function to each pixel individually is computationally expensive. Therefore, in this work, we utilize a simpler approximation which is computationally cheaper, yet delivers high accuracy in variance computation. We utilize the characteristics of a standard Gaussian curve, $g(c) = \frac{1}{\sigma_r\sqrt{2\pi}} e^{-0.5\left(\frac{c}{\sigma_r}\right)^2}$ evaluated on its variance value that yields $g(c = \sigma_r) \approx 0.6 g_{max}$, where $g_{max}$ is the maximum value of the Gaussian function. Note that, under the assumption that the predictive probability distribution $x'_r$ is Gaussian, we can simply capture its maximum as $\max(x'_r)$. To reduce the impact of inaccuracies that may arise from measuring the maximum value, it is possible to calculate a 3-point average of consecutive neighboring points around the maximum value. Then, the variance of $x'_r$ can be estimated by counting the number of elements in $x'_r$ that is greater than $0.6 \max(x'_r)$. By further normalizing this quantity, it is possible to obtain an accurate estimate for Eq. 10. We followed this procedure to estimate uncertainty in our proposed PixCUE framework.

### 3.5 Model implementation

The underlying DL model of the PixCUE framework was trained using the Pytorch framework on NVIDIA V100 GPU. We utilized a total of six iterations in our implementation of the VN. Rectified Adam optimizer with a learning rate of 0.0001 was utilized and the model with the best validation loss was selected as the final model. The k-space data was undersampled using an equidistant undersampling pattern with 8% of center lines always acquired. A zero-filled image reconstructed from the undersampled data was utilized as an input to the network. The dynamic range of the target image (reference fully sampled) was quantized to 8-bit (i.e., 256-pixel intensity levels) to form the target distribution at a given pixel location. The categorical cross-entropy loss (Eq. 6) was utilized for training. At the time of inference, the predicted probabilities were utilized to compute the final image using Eq. 7 and the uncertainty map using Eq. 10.

## 4 Experiments

### 4.1 Dataset

The dataset utilized for the training was the fastMRI multi-coil brain dataset [71] which consisted of T1, T2, T1-POST contrast, and FLAIR acquisitions in k-space. The distribution of the dataset for different contrasts is provided in Table 1. All the experiments were performed on the validation dataset.

Table 1: Contrast distribution of the number of multi-slice images utilized for training and validation.

| Contrast | Training | Validation |
|---|---|---|
| FLAIR | 343 | 107 |
| T1 | 498 | 169 |
| T1-POST | 949 | 287 |
| T2 | 2678 | 815 |
| **Total** | **4468** | **1378** |

## 4.2 List of Experiments

We conducted several experiments to assess the performance of PixCUE and visualized the local correspondence of the estimated uncertainties with the actual absolute error in the reconstruction. Experiments I to IV are designed likewise with various adversarial conditions introduced at the input. Experiment V assesses the local correspondence of PixCUE with the MC Drop-Out method. Further, in Experiment VI, we formulate an empirical relationship between the estimated uncertainties using PixCUE and the reconstruction performance metrics. The details of our experiments are listed below.

*Experiment I – k-space undersampling:* In this experiment, we evaluated the performance of PixCUE for accelerated MRI reconstruction. Specifically, k-space was undersampled in the phase encoding direction by a factor of four using a random undersampling pattern with the central 8% of phase encode lines always sampled. Both image reconstruction and uncertainty estimation were then performed using the undersampled k-space. Reconstruction errors and uncertainty maps were compared.

*Experiment II - SNR variation:* Imaging parameters such as TE/TR/flip-angle/image resolution result in different signal-to-noise (SNR) ratios in MR images. In this experiment, we simulated SNR variations by adding Gaussian noise to the complex-valued k-space and evaluated the effect on the reconstruction and how the estimated uncertainty captured out-of-distribution information introduced by the noise in the input datasets.

*Experiment III - Sampling pattern and acceleration factor variation:* DL models are often trained for a certain undersampling pattern, however in practice undersampling patterns may vary due to changes in the imaging field of view and k-space sampling. This experiment analyzed how the estimated uncertainty varies if the sampling pattern is changed. The model was trained with an undersampling factor of four, but during the image reconstruction, the undersampling factor was changed to six with the central 6% of phase encode lines sampled.

*Experiment IV - Pathology*: In this experiment, we evaluated PixCUE on a case consisting of pathology (i.e., tumor) for each contrast. We investigated how the estimated uncertainty captured out-of-distribution information introduced by pathology. The reconstruction error and the estimated uncertainty were calculated and compared.

*Experiment V – Comparison between PixCUE with the MC Drop-Out method.* To compare the uncertainty estimations using PixCUE and MC methods, we performed an MC Drop-Out simulation experiment using the VN backbone network which yields multiple predicted distributions. We then computed an average probability distribution as a representation of the MC Drop-Out inferences and calculated the corresponding variance similar to PixCUE. Then we visualize the uncertainty maps of these two methods. We also plot the joint distribution of the uncertainty values predicted by these two approaches for randomly selected 100 pixels of an image in order to investigate the correlation between the two approaches. In the MC experiment, we utilized a drop-out fraction of 0.2 and performed 50 iterations for the variational inference.

*Experiment VI - Uncertainty vs quantitative metrics:* In this experiment, we empirically evaluated the relationship between the estimated uncertainty using PixCUE and qualitative metrics including normalized mean squares error (NMSE), peak signal-to-noise ratio (PSNR), and structural similarity (SSIM) index. We calculated the average uncertainty and quantitative metrics on all the images in the validation dataset and performed curve fitting using linear regression and the $R^2$ value was utilized to measure the goodness of fit.

## 5 Results

Fig. 2 shows the reconstructed images along with the computed uncertainty and the absolute error images for four different contrasts. Visual inspection showed that the uncertainty estimated by PixCUE correlates with the error suggesting that uncertainty maps can be utilized as a proxy for error maps. Noticeably, a hallucination (blue arrow) in the reconstructed image is estimated as a region of high uncertainty. When noise was added to the images to simulate different SNR scenarios (Fig. 3), minor changes were observed in uncertainty. When the sampling pattern was changed (Fig. 4) to an undersampling factor of six, still the pixel classification framework was able to capture features that showed greater error as seen by the substantial spatial correlation between uncertainty and the absolute error. Fig. 5 shows the images with pathology, it can be observed that the uncertainty is higher in the regions of the tumor. Since the tumor data is underrepresented during training, the PixCUE framework was able to identify it by predicting high spatial uncertainty in the tumor region.

Fig. 6 demonstrates the correlation between PixCUE and the MC method. By comparing the left and right columns of Fig. 6(a), we observed that PixCUE produced almost similar uncertainty maps to that of the MC method. Further, when the joint distribution of uncertainties of PixCUE and MC method was visualized, a high linear correlation was observed and the marginal distributions of the two methods showed similar profiles. This shows that the pixel classification framework with a single inference captures uncertainty up to a similar extent that is captured by the MC method but with a less computational cost.

Fig. 7 shows the plot of quantitative scores with respect to the computed average uncertainty of the image along with the equation of the fitted curve and $R^2$ value (goodness of fit). We utilized linear regression curve fitting. We observed that a linear fitting curve provided the best fit for NMSE (Fig. 7 Top row) and SSIM (Fig. 7 Bottom row), while an exponential curve fitted better for the PSNR (Fig. 7 Middle row). The $R^2$ value was substantially higher for NMSE ($\geq 0.66$). In terms of SSIM, the $R^2$ value is higher for only T2 contrast. The high $R^2$ values of uncertainty for NMSE demonstrate that the uncertainty estimated by PixCUE is a good proxy metric for estimating the actual error of the reconstructed images. These plots visually demonstrate an empirical relationship between the quantitative metrics of the reconstructed images and the predicted uncertainty, which is crucial for the design of accurate image reconstruction models.

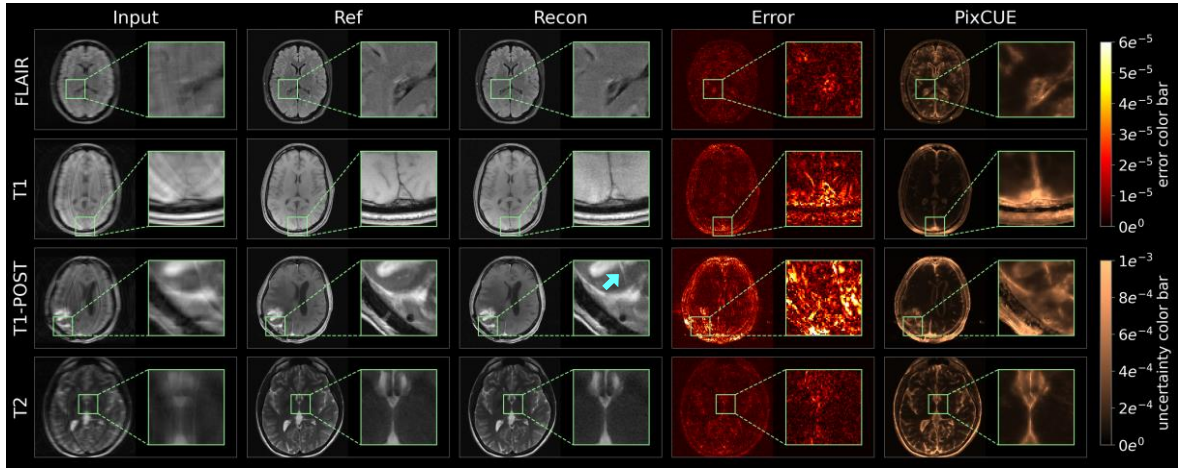

**Fig. 2:** Results for Experiment I. *Input*: zero-filled input image to DL network reconstructed from undersampled k-space; *Ref*: ground truth fully sampled image; *Recon*: reconstructed image using DL pixel classification network; *Error*: absolute error image; *PixCUE*: uncertainty obtained using the pixel classification framework. Visual inspection suggests that there exists a correlation between the error images and calculated uncertainties. The blue arrow shows a region of hallucination (artificial feature) with high uncertainty.

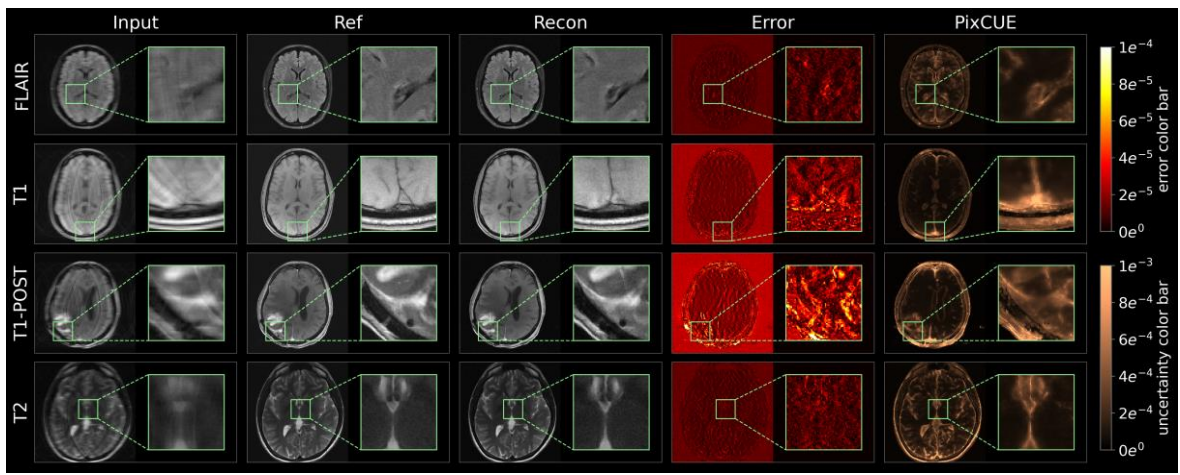

**Fig. 3:** Results for Experiment II. *Input*: noisy zero-filled input image to DL network reconstructed from undersampled k-space; *Ref*: ground truth fully sampled image; *Recon*: reconstructed image using DL pixel classification network; *Error*: absolute error image; *PixCUE*: uncertainty obtained using the pixel classification framework. Visual inspection suggests that estimated uncertainty is sensitive to noise as evident from uniform higher values within the brain region.

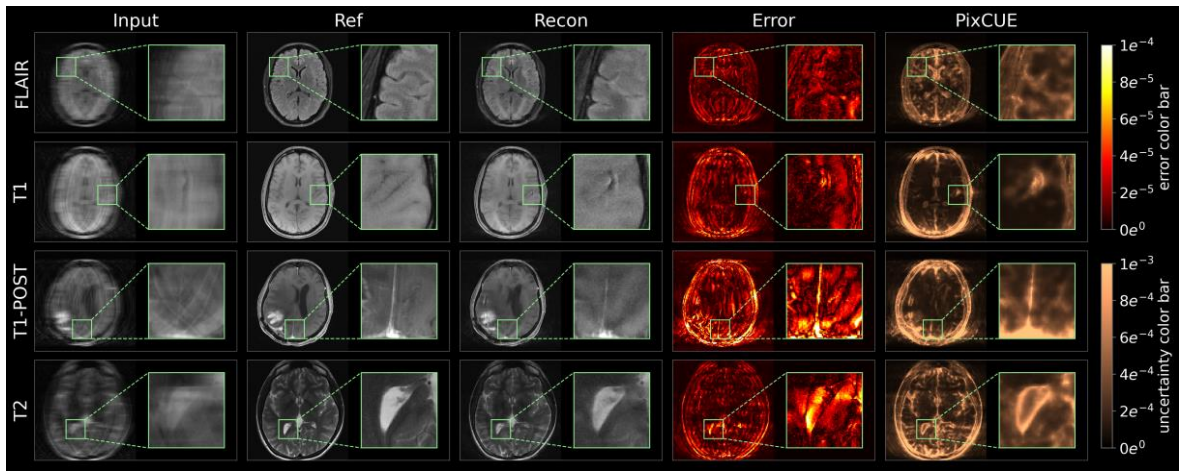

**Fig. 4:** Results for Experiment III. *input*: zero-filled input image to DL network reconstructed from undersampled k-space (undersampling factor of 6); *Ref*: ground truth fully sampled image; Recon: reconstructed image using DL pixel classification network; *Error*: absolute error image; *PixCUE:* uncertainty obtained using the pixel classification framework. It can be observed that as the undersampling pattern diverges from the training undersampling pattern, the algorithm suffers marked degradation in performance. The pixel classification framework was able to capture this divergence of the sampling pattern.

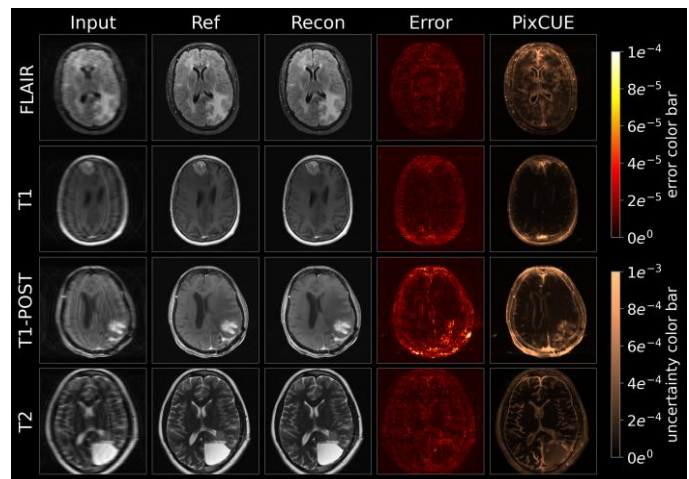

**Fig. 5:** Results for Experiment IV. *input*: zero-filled input image to DL network reconstructed from undersampled k-space (undersampling factor of 6); *Ref*: ground truth fully sampled image; Recon: reconstructed image using DL pixel classification network; *Error*: absolute error image; *PixCUE:* uncertainty obtained using the pixel classification framework. The pixel classification uncertainty estimate was sensitive to underrepresented data (tumor).

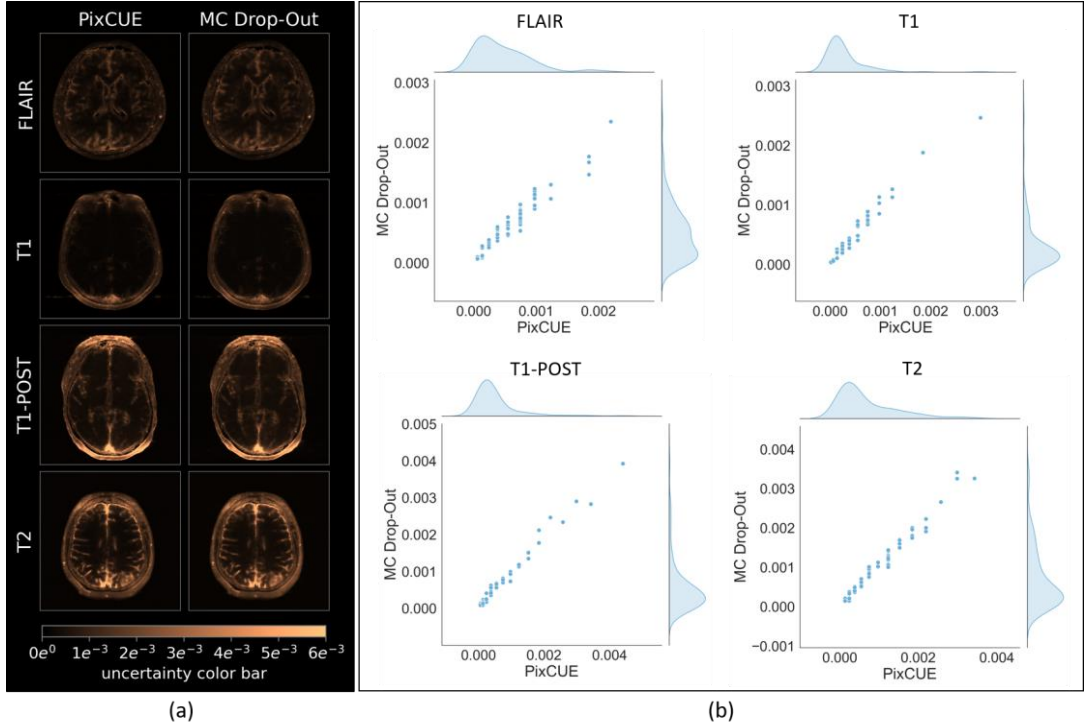

**Fig. 6:** Results for Experiment V. (a) *PixCUE:* uncertainty obtained using the pixel classification framework. *MC Drop-Out:* uncertainty obtained using the MC method. (b) Joint distribution of the uncertainty values predicted by the proposed PixCUE method and the MC method utilizing 100 randomly selected pixels from each uncertainty map shown in Fig. 6(a).

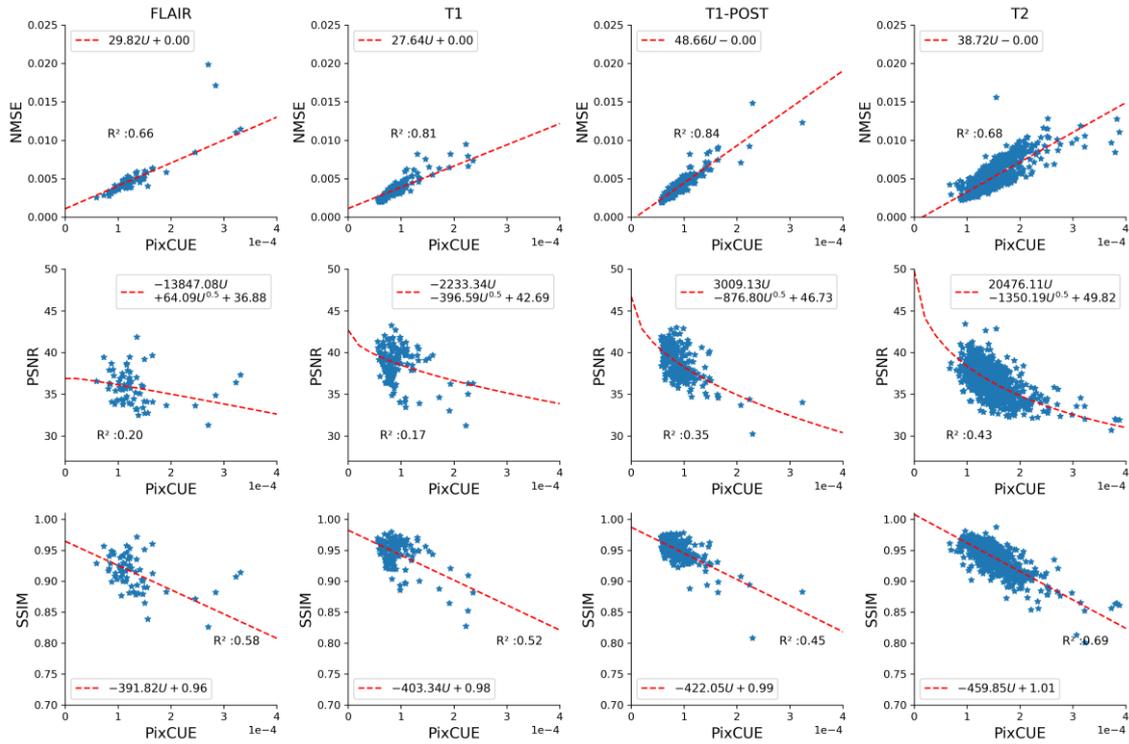

**Fig. 7:** Results for Experiment VI. Relationship between quantitative metrics (NMSE: top row, PSNR: middle row, SSIM: bottom row) and uncertainty estimated by PixCUE; legend shows the equation of the fitted curve, and the $R^2$ value represents the goodness of fit (higher the better fit).

## 6   Discussion

Our experiments show that the proposed PixCUE framework can produce reconstructed images along with the uncertainty estimations at no extra computation burden. The predicted uncertainty was found consistent with the previous literature [37], [38]. Our experiments highlighted that PixCUE is capable of capturing the actual image error in the reconstructed image. PixCUE is also robust to practical adversarial scenarios such as noise, change of sampling pattern, and identification of the out-of-distribution (underrepresented) data such as pathologies in images. However, contrast changes such as pathology do not necessarily degrade the reconstruction image quality. Even with the underrepresented pathology data, the model was able to reconstruct the tumor region faithfully provided sufficient training, and the reconstruction error is captured better by our PixCUE framework. Our overall results suggest modeling uncertainties is a critical step for accurate DL-based image reconstruction with increased explanatory power in their reconstruction errors.

The computational cost is also an important aspect of uncertainty modeling that needs to be considered for practical purposes. In general, uncertainties are estimated using variational inference which requires multiple inferences of the same input. Literature suggests around 50 inferences implying 50 times more computation [35], [38] in addition to image reconstruction time. The increased computational time can make real-time uncertainty estimation challenging. On the other hand, our PixCUE framework does not require additional computation and the uncertainty emerges as a co-product of the standard image reconstruction. For instance, the uncertainty can be computed in the order of a few seconds, whereas MC methods can take minutes for the same image which may not meet the demand for real-time applications. Nevertheless, Experiment V shows a close correlation between the uncertainty estimations produced by PixCUE and the MC method. The real-time uncertainty estimation capability of PixCUE provides an opportunity to design future uncertainty-guided image reconstruction methods.

In this work, we included several practical scenarios expected in clinical practice. However, we have some limitations of the study, for instance, we simulated the most frequently expected out-of-distribution cases including noise, and change of sampling patterns but other out-of-distribution cases including hardware configuration and instrument-related changes have not been explored. Out-of-distribution can also occur when dealing with variations in anatomical structures. For example, we can expect that a model trained on brain datasets of a particular contrast would have reduced performance when applied to whole-body imaging.

It is also worthwhile to note that, in this work, we have employed a state-of-the-art network with reconstruction performance similar to the fastMRI challenge results, indicating the model parameters are well-fitted. However, we have not conducted an extensive test on different network architectures and/or with over- or under-fitted model parameters. These variations can further influence the quantitative relationship between uncertainties and reconstruction errors which warrants further investigations, and the proposed framework can be readily applied to such studies.

## 7   Conclusion

In this work, we estimated and evaluated uncertainties during MR image reconstruction using a deep pixel classification approach. A novel method of estimating uncertainty, PixCUE was proposed and validated on large datasets with four different contrasts. Different realistic practical scenarios were simulated and their impacts on uncertainty were compared. Overall, it was observed that the uncertainty estimated by PixCUE corresponds well with the actual error in image reconstruction, hence can be utilized as a proxy for error. The lower computational cost of calculating uncertainty makes it practical for time-constrained medical image reconstruction applications. Also, an empirical relationship between uncertainties and quantitative image quality metrics is identified in this paper and can be useful in estimating image reconstruction errors in practice when dealing with noise and sampling pattern changes.

## Acknowledgments

This work was conducted as a part of the projects titled "Simultaneous to synergistic MR-PET: integrative brain imaging technologies" funded by the Australian Research Council Linkage Program (LP170100494) and "Biophysics-informed deep learning framework for magnetic resonance imaging" funded by the Australian Research Council Discovery Program (DP210101863).